%% file: paper.tex
\documentclass[sigconf,authorversion,nonacm,10pt]{acmart}

\usepackage{enumitem}
\usepackage{multirow}
\usepackage[labelfont=bf,textfont=md,skip=3pt,font=footnotesize]{caption}
\usepackage{subcaption}
\usepackage{balance}

\newcommand{\name} {PanoRadar}
\newcommand{\header}[1]{\vskip 0.1cm \noindent{\bf #1}}

\newcommand{\figref}[1]{Fig.~\ref{#1}}
\newcommand{\tabref}[1]{Tab.~\ref{#1}}
\newcommand{\secref}[1]{\S~\ref{#1}}

\newcommand{\eqnref}[1]{Eqn.~(\ref{#1})}

\begin{document}
\title{Enabling Visual Recognition at Radio Frequency}

\author{Haowen Lai}
\affiliation{
  \institution{University of Pennsylvania}
}
\email{hwlai@seas.upenn.edu}
\author{Gaoxiang Luo}
\affiliation{
  \institution{University of Pennsylvania}
}
\email{luo00042@seas.upenn.edu}
\author{Yifei Liu}
\affiliation{
  \institution{University of Pennsylvania}
}
\email{freddyl6@seas.upenn.edu}
\author{Mingmin Zhao}
\affiliation{
  \institution{University of Pennsylvania}
}
\email{mingminz@seas.upenn.edu}

\begin{abstract}
\input{src/abstract-3.tex}
\end{abstract}


\settopmatter{printfolios=true}

\maketitle

\input{src/intro-6.tex}
\input{src/related-work-3.tex}

\input{src/overview-2.tex}
\input{src/static-3.tex}

\input{src/motion-estimation-5.tex}

\input{src/3d-ML-4.tex}
\input{src/implementation.tex}

\input{src/evaluation-3.tex}

\input{src/limitations.tex}

\input{src/conclusion.tex}

\begin{acks}
We are grateful to Xin Yang, Zitong Lan, Dongyin Hu, Ahhyun Yuh, and Zhiwei Zheng for their feedback.
We also thank Yiqiao Liao for his contributions during the early development of this project.
\end{acks}

\input{src/appendix}

\bibliographystyle{ACM-Reference-Format}
\bibliography{references}

\end{document}

%% file: src/abstract-3.tex
This paper introduces \name{}, a novel RF imaging system that brings RF resolution close to that of LiDAR, while providing resilience against conditions challenging for optical signals. Our LiDAR-comparable 3D imaging results enable, for the first time, a variety of visual recognition tasks at radio frequency, including surface normal estimation, semantic segmentation, and object detection. \name{} utilizes a rotating single-chip mmWave radar, along with a combination of novel signal processing and machine learning algorithms, to create high-resolution 3D images of the surroundings. Our system accurately estimates robot motion, allowing for coherent imaging through a dense grid of synthetic antennas. It also exploits the high azimuth resolution to enhance elevation resolution using learning-based methods. Furthermore, \name{} tackles 3D learning via 2D convolutions and addresses challenges due to the unique characteristics of RF signals. Our results demonstrate \name{}'s robust performance across 12 buildings.

%% file: src/intro-6.tex
\section{Introduction}

\begin{figure}[t]
    \centering
    \includegraphics[width=\linewidth]{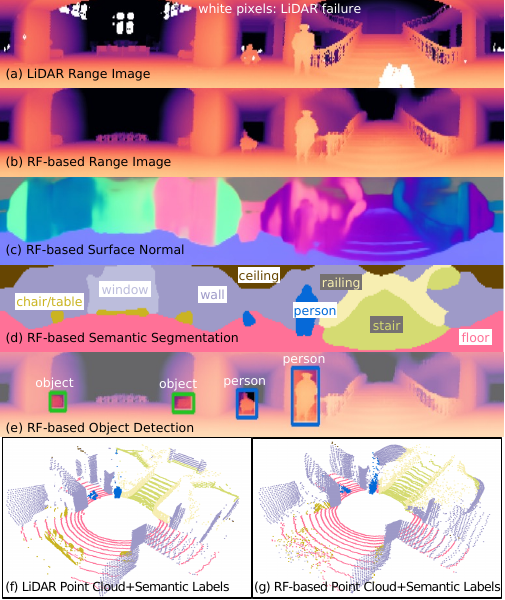}
    \vspace{-10pt}
    \caption{RF imaging and visual recognition with \name{}. This figure illustrates the capabilities of our system, showing (a) the 3D panoramic LiDAR range image as a reference, and (b) the RF-based prediction generated by our system. 
    Our LiDAR-comparable results enable a variety of visual recognition tasks, including (c) surface normal estimation, (d) semantic segmentation, and (e) object detection and human localization. Additionally, we present (f) the LiDAR 3D point cloud color-coded with manually-annotated semantic labels, and (g) the predicted RF-based point cloud color-coded by the corresponding predicted semantic categories, which offers an enriched understanding of the 3D surroundings.}
    \label{fig:teaser}
    \vspace{-10pt}
\end{figure}

The emergence of robotic and autonomous systems in areas such as transportation, search and rescue, construction, healthcare assistance, and warehouse management is poised to improve efficiency, safety, and human well-being across a diverse range of sectors~\cite{habib2010robot, takahashi2010developing, dhiman2022fire}. To ensure accurate and robust perception of the surroundings, Radio Frequency (RF) signals-based sensing and imaging~\cite{Pointillism, HawkEye, radatron, satat2018towards} have appeared as promising techniques. These RF systems offer distinct advantages over traditional optical sensors, with resilience against environmental challenges such as dust, fog, smoke, and adverse lighting conditions~\cite{see_through_smoke, CMU_mmWave_pc}.

The fundamental limitation of RF sensors compared to optical ones, however, lies in their poor resolution~\cite{fang2020superrf,RF-pose, 3drimr}. Unlike cameras, where millions of pixels can be integrated into a CMOS sensor, RF sensors typically consist of significantly fewer antennas. This constraint results in limited angular resolution, causing objects in close proximity to appear smeared or indistinct~\cite{radatron}. Such limitations have led to substantial challenges in capturing high-resolution RF images that convey object and environment details.

Past research has explored different methods to enhance the resolution of RF images.
Some techniques focus on specific categories such as humans~\cite{RF-pose, RF-pose3d,witrack2, Wideo} or vehicles~\cite{radatron, HawkEye, Pointillism, 3drimr}, leveraging category-specific prior knowledge or generative models. 
Other solutions employ synthetic aperture radar techniques by moving radars on slide rails~\cite{yanik2020development, yanik2019near, HawkEye}; however, this method is unsuitable for mobile robots due to the cumbersome rail sizes (e.g., 1.2 m~\cite{yanik2020development}) and slow scanning speed (e.g., 5 mins~\cite{HawkEye}). 
Additionally, researchers have sought to improve resolution by opportunistically leveraging the external motion from the robot~\cite{UCSD_3D_pc, CMU_mmWave_pc}.
This approach, however, can only enhance resolution in the moving direction and becomes ineffective when the robot is static.

This paper introduces \name{}, an RF imaging system that enhances sensing resolution to a level similar to LiDAR, enabling a wide array of visual recognition tasks with RF signals. \name{} operates by rotating a single-chip mmWave radar with a motor, forming a dense cylindrical array of antennas. Our system then utilizes a combination of novel signal processing and machine learning algorithms to create high-resolution 3D images of the surroundings. Fig. 1 shows an example output from our system and compares it with a panoramic 3D LiDAR (Ouster OS0-64, 64-beam, $\sim$\$9000 USD). Fig. 1(a) presents the range image captured by the LiDAR, while Fig. 1(b) showcases \name{}'s output based solely on RF signals. Our predictions offer a resolution comparable to 3D LiDAR, capturing detailed structures of a building, including walls, floors, ceilings, stairs, as well as humans and objects (e.g., chairs, benches). Equipped with this LiDAR-comparable range image, \name{} enables visual recognition tasks with RF signals, such as surface normal estimation, semantic segmentation, and object and human detection (Fig. 1(c)-(e)). 
Outputs from \name{} can be viewed in 3D (Fig. 1(g)), with predicted 3D point clouds color-coded by semantic predictions, offering an enhanced visualization and understanding of the surroundings. Fig. 1(f) provides a reference captured by LiDAR with manual annotations.

\begin{figure}[t]
    \centering
    \includegraphics[width=0.7\linewidth]{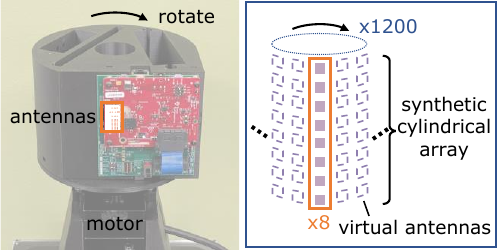}
    \caption{\name{} design. Left: our system rotates a single-chip mmWave radar using a motor, with its linear antenna array placed vertically. Right: this rotation emulates a dense cylindrical array of antennas.}
    \label{fig:system_illustration}
    \vspace{-13pt}
\end{figure}

A key design of \name{} involves the rotation of a mmWave radar. As illustrated in Fig.~\ref{fig:system_illustration}, our system employs a commercial off-the-shelf (COTS) single-chip mmWave radar and rotates it around the vertical axis using a motor. Since the linear array is placed vertically, this rotation forms a dense cylindrical array of synthetic antennas (8x1200). 
This unique design offers several distinct benefits: 1) The rotation around the vertical axis greatly enhances the azimuth resolution to 2.6 degrees (\secref{sec:static}). 2) By vertically placing the linear array, our system can perform beamforming along the vertical axes, enabling 3D imaging of the environment. 3) The rotation also expands the otherwise static limited field of view (FOV), typically ranging from 30-60 degrees~\cite{awr1843beamwidth}, providing panoramic sensing of the environment. 4) With COTS radar and motor and a compact rotation radius of 8 cm, the system ensures low cost, fast capture time and enhanced mobility. Our design stands apart from existing mechanical radars~\cite{burnett2022boreas, oxford_Radar, prob_unknown, dont_need_compen}, which rotate expensive and customized directional antennas and are limited to 2D mapping.

Despite the above benefits, designing and implementing \name{} presents multiple challenges. The first one is the external motion during sensor rotation (e.g., the sensor on a moving robot). Coherent combination of all the synthetic antennas requires their precise location at a sub-wavelength level (e.g., $\lambda/2=1.9$ mm at 79 GHz). Unfortunately, commonly used sensors like IMUs or wheel odometers cannot track robot motion with such accuracy~\cite{IMU_wheel_odom, UCSD_3D_pc}. Ideally, we would leverage reflected RF signals from the environment for motion estimation (e.g., the Doppler effect). However, this is difficult as the Doppler effect measures the radial speed along the direction of the reflector, and without knowing the exact AoA of reflectors, we cannot recover the velocity (i.e., speed and heading direction). What is worse, AoA and Doppler effect are intertwined, as they both manifest as the frequency shift across multiple antenna measurements.~To overcome these challenges, we design novel signal processing algorithms to untangle this ambiguity between AoA and Doppler effect. Our algorithm tracks reflections from the same reflector during radar rotation and leverages the directional characteristic of mmWave antennas to identify its AoA (i.e., the same reflector would have the strongest reflection when it's at 90 degrees with respect to the radar). By incorporating multiple reflectors from the environment and observing radial speed from different angles, \name{} achieves robust motion estimation and compensation.

The second challenge pertains to the limited elevation resolution.  While our system achieves high resolution along the azimuth (1200 virtual antennas) and range (3.75cm with 4GHz bandwidth) dimensions, its elevation resolution is limited due to significantly fewer antennas along the vertical axis (8 for the COTS radar used). To overcome this, we leverage the high azimuth and range resolution to enhance the elevation resolution, given that spatial dimensions are not independent in 3D environments. While direct information transfer between these dimensions is generally impractical, it becomes feasible in our case due to the inherent low-rank structures and unique attributes of indoor environments~\cite{sparsity,depthmc,mcnoise}. For example, common indoor structures like walls, floors, ceilings, stairs, and furniture introduce regularities in range images. Consistent depth cues from these surfaces can be used to infer information vertically. 
On top of that, ML models (e.g., CNN) can effectively manage these cross-dimensional dependencies, using interpolation to construct continuous surfaces from sparse observations. ML models can also learn to deal with radar artifacts~\cite{radatron}, such as performing deconvolution to counteract energy spreading in beamforming algorithms. This capability is beneficial given that we have already possessed high azimuth and range resolution, minimizing the number of reflectors to be resolved along the elevation dimension. Hence, we employ learning-based methods to capitalize on these insights, i.e., models trained with a large dataset of paired LiDAR and RF data to enhance elevation resolution for RF inputs. By recognizing and utilizing consistent patterns in indoor scenes, our model allows for more accurate predictions and a richer understanding of the 3D environment.

Finally, the design of ML models for high-resolution RF imaging and the subsequent visual recognition tasks presents a set of challenges due to the unique characteristics of RF signals and the complexity of learning with 3D panoramic data. At the forefront of these challenges is 3D learning. While 3D convolution might seem an intuitive choice for learning the 3D structure from our 3D RF data, it rapidly becomes impractical. Consider voxelizing a space of 20m x 20m x 5m with 2cm cubes; the resulting tensor size of 1000x1000x250, not even accounting for the channel dimension, would require excessive processing and memory with 3D CNN. This is compounded by the challenge of learning from highly-imbalanced occupancy grids (i.e., less than 1\% of the voxels would be filled). 
Our approach utilizes 2D models to facilitate 3D learning by taking advantage of the intrinsic sparsity in both LiDAR and RF data. First, we repurpose the range dimension as channels and feed the 3D RF data into 2D models. Unlike conventional CNNs, we begin our models with an 4x reduction in the channel dimension, optimizing for efficiency given the sparse RF reflections along the range dimension. Second, our model is designed to predict a 2D range image that indicates the distance to the first reflector from various directions, essentially converting it into a 2D prediction task.
In addition to this challenge, our design also tackles other issues such as: 1) the inability of LiDAR to capture glass and the resulting incorrect range supervision for RF inputs during training; 2) multi-path reflections in RF signals, which could result in ghost objects;
and 3) the need for our model to harness to the panoramic nature of RF imagery.

We build a prototype of \name{} and deploy it on a mobile robot with a 3D LiDAR for reference. We also annotate LiDAR points with semantic and object labels as ground truth for the corresponding recognition tasks. We conduct experiments in 12 different buildings around our campus and run leave-one-building-out training/testing splits to ensure generalization across buildings. For motion estimation, \name{}’s mean speed error is 8.48 mm/s, and the mean heading direction error is 1.09$^\circ$. For 3D imaging, the mean error of our range prediction is 15.76 cm. \name{} achieves a 8.83$^\circ$ surface normal estimation error, a mean intersection over union (mIoU) of 48.00 for semantic segmentation, and an average precision ($\mathrm{AP^{30}}$) of 52.34\% for object detection.

The key contributions of the paper are as follows:
\begin{itemize}[leftmargin=*,itemsep=0pt,topsep=0pt]
    \item We introduce the first RF imaging system that achieves resolution close to that of LiDAR. It enables, for the first time, visual recognition with RF, including surface normal estimation, semantic segmentation, and object detection.
    \item We present a novel design that integrates a COTS mmWave radar with a motor to significantly enhance sensing resolution and FoV, while ensuring that the system remains compact, low-cost, and practical for mobile robots.
    \item We propose a novel robot motion estimation algorithm that accurately estimates and compensates for robot motion, allowing for coherent combination of radar signals.
    \item We introduce a learning model that effectively enhances vertical resolution using high azimuth and range resolutions, while maintaining efficiency with 2D convolutions.
    \item We evaluate our system across 12 diverse buildings, demonstrating its feasibility, accuracy, and robustness in various environments. We will release code and dataset to facilitate future research in this direction.
\end{itemize}

%% file: src/related-work-3.tex
\section{Related Work}
\header{RF Imaging and Sensing:}
Recent years have witnessed an increasing interest in wireless and RF imaging. Various techniques have been explored including WiFi~\cite{wiffract, dloc, P2SLAM}, RFID~\cite{tagscan, rfusion, rfid_finger}, radars~\cite{CMU_mmWave_pc, UCSD_3D_pc, yanik2019near, yanik2020development, osprey, adib20143d, witrack2, rfsleep, medicationselfassessment}, and combinations of multiple sensors~\cite{lu2020milliego, see_through_smoke} to perceive and image the object.
Synthetic aperture radar (SAR)~\cite{SAR_basis0,SAR_basis1,SAR_basis2,SAR_basis3} is a common technique used to enhance imaging resolution.
Previous work~\cite{wiffract, fang2020superrf, yanik2019near, yanik2020development, HawkEye, 3drimr, mask_not_matter} uses horizontal and vertical sliders to move the radar, emulating a planar array.
However, this method is limited by its long scanning time (e.g., 5 minutes in \cite{HawkEye}) and cumbersome setup.
Circular SAR~\cite{cir_basis0, cir_AF} has also been explored with enhanced efficiency and panoramic sensing capability. Despite its improvements, it has mainly been adopted in large static setups, such as security scanners~\cite{airport_scanner} and CT scanners~\cite{CT_scanner}, as coherent imaging is challenging in the presence of motion.
There is also previous work that utilizes 360-degree radars for autonomous vehicles~\cite{burnett2022boreas, oxford_Radar, prob_unknown, dont_need_compen}.
These systems use mechanical radars, e.g., Navtech CTS350-X~\cite{Navtech_spec}, which are expensive and only provide 2D mapping of the surroundings.

Machine learning is another avenue to enhance sensing resolution.
2D CNNs have been used in prior systems~\cite{CMU_mmWave_pc, radatron, see_through_smoke}. 
However, these methods are limited to 2D imaging, which are not sufficient for tasks that require 3D understanding of the surroundings.
Some work also utilizes 3D convolutions for 3D imaging~\cite{HawkEye, 3drimr, deeppoint, mmMesh}. 
However, 3D convolutions are computationally heavy and are prone to overfitting when learning from sparse radar reflections. 
On top of that, learning-based imaging and sensing have primarily focused on humans and vehicles. 
For example, learning-based human sensing includes pose estimation~\cite{m3track, wipose, RF-pose, RF-pose3d, mmMesh, m4esh}, action recognition~\cite{RF-action}, etc., while learning-based vehicle perception focuses on detection~\cite{Pointillism, radatron} and silhouette reconstruction~\cite{HawkEye, 3drimr}.
However, focusing on specific classes ignore other categories that are equally important in various contexts. Furthermore, these algorithms/models integrate category-specific priors, which limits their applicability as a general imaging solution.
There are also learning based approaches to extract semantic information from surrounding objects using RF \cite{see_through_smoke}.
However, they have constraints on the radar's position, and only classifies objects given a predefined region. These factors limit its application in more complex imaging tasks such as object detection.

\header{Computer Vision:}
Significant efforts have driven substantial advancements in the field of visual recognition based on cameras~\cite{DeepLabv3, DeepLabv3+, rcnn, fast-rcnn, faster-rcnn} and LiDAR~\cite{PointNet, PointNet++, voxnet, rangenet++}, as well as their fusion~\cite{adafusion, deepfusion}.
Cameras excel at capturing high-resolution spatial information in the form of images,
yet they rely heavily on adequate lighting for imaging, rendering them unusable in low-light or harsh weather conditions.
Meanwhile, LiDAR sensors utilize lasers to capture environmental geometry, producing 3D point clouds. However, they are susceptible to airborne particles like smoke and dust~\cite{LiDAR_fail1, CMU_mmWave_pc}, and 3D LiDAR systems are expensive.

%% file: src/overview-2.tex
\section{Overview}

\begin{figure}[htbp]
    \centering
    \includegraphics[width=\linewidth]{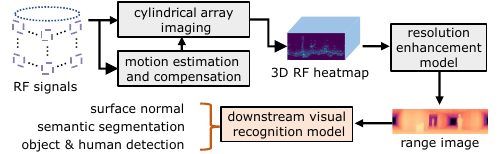}
    \caption{\name{} architecture with four components: a 3D imaging system for cylindrical arrays (\secref{sec:static}), a motion estimation and compensation algorithm (\secref{sec:motion_estimation_and_imaging}), a resolution enhancement and range image estimation model (\secref{sec:range_estimation_ml}), and visual recognition heads for downstream tasks (\secref{sec:visual_recog_model}).}
    \label{fig:overview}
    \vspace{-10pt}
\end{figure}

\name{} is an RF imaging system that delivers LiDAR-comparable resolution and enables visual recognition.
\figref{fig:overview} illustrates the system architecture of \name{}, consisting of four components: 
1) 3D imaging with a rotating radar: we describe cylindrical array imaging and analyze imaging resolution (\secref{sec:static});
2) Motion estimation and compensation: we present our novel algorithms for accurate estimation of robot motion, and efficient imaging algorithms with compensation (\secref{sec:motion_estimation_and_imaging});
3) Vertical resolution enhancement and range image estimation: we detail our design of ML methods that efficiently learn 3D structures with 2D models (\secref{sec:range_estimation_ml}); and 
4) Visual recognition heads: we outline our design of models for various downstream recognition tasks (\secref{sec:visual_recog_model}).

%% file: src/static-3.tex
\section{Cylindrical Array Imaging}
\label{sec:static}

We begin with the fundamentals of RF imaging using a rotating radar. In this section, we focus on a specific scenario in which the robot platform remains stationary --i.e., the only movement experienced by the radar is the motor-driven rotation.
Recall from \figref{fig:system_illustration} that our system employs a unique design that uses rotation to create a synthetic cylindrical array. Specifically, assuming our system has $A$ antennas placed vertically, each with a height of $h^a, a=1,…,A$, and the antennas are rotating with raduis $r$ and angular speed $\omega$, the location of antenna $a$ at time $t$ can be expressed as:
\begin{equation}
\vec{p}^a_t = (r\cos(\omega t), r\sin(\omega t), h^a).
\end{equation}

Given a cylindrical array, consider an imaging direction of interest $\vec{d}$ with azimuth $\theta_d$ and elevation angle $\phi_d$:
\begin{equation}
    \vec{d} = (\cos\phi_d \cos\theta_d, \cos\phi_d \sin\theta_d, \sin\phi_d).
\end{equation}
The key to forming a narrow beam along $\vec{d}$ is to coherently combine all the antennas by compensating for their difference in the distance to the plane perpendicular to $\vec{d}$~\cite{radar_fundamentals}.
Since this distance is the projection of $\vec{p}^a_t$ to $\vec{d}$ (equivalently $\vec{d} {\cdot} \vec{p}^a_t$ since $\vec{d}$ is a unit vector), our 3D beamforming can be achieved via:
\begin{equation}
    \boldsymbol{B}(\vec{d}) = \sum_{a,t} \boldsymbol{S}^a_t 
        \exp \left(j2\pi \frac{2\vec{d} {\cdot} \vec{p}^a_t}{\lambda} \right),
\label{eq:static_imaging}
\end{equation}
where $\boldsymbol{S}^a_t$ represents the intermediate frequency (IF) signals from antenna $a$ at time $t$, $\boldsymbol{B}(\vec{d})$ is the beam formed, and $2\vec{d} {\cdot} \vec{p}^a_t$ is attributed to the round-trip phase difference.
We perform range FFT on $\boldsymbol{B}(\vec{d})$ to obtain the range dimension, indicating the distance of objects along this direction.
By querying beams along a 2D grid of azimuth and elevation angles, we obtain a 3D (i.e., azimuth x elevation x range of size 512 x 64 x 256) tensor showing reflections from the 3D surroundings.
We note that since the mmWave radar antennas used here are not omni-directional (i.e., they have a 3-dB beamwidth of around 30$^\circ$ along the azimuth dimension), we limit the antenna positions used for summation in \eqref{eq:static_imaging} to those centered around $\theta_d$ for efficiency purposes.

\begin{figure}[t]
    \centering
    \includegraphics[width=\linewidth]{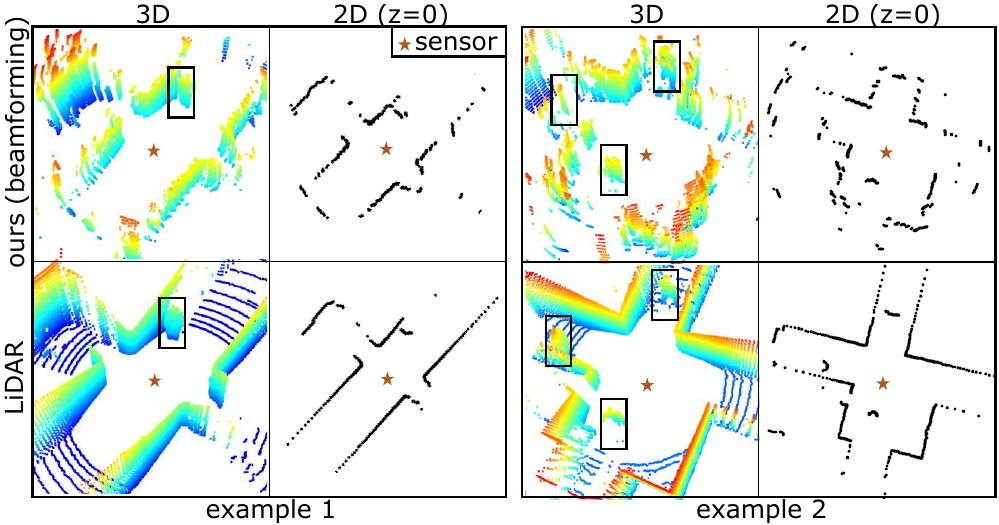}
    \caption{RF imaging results with a stationary robot. Our beamforming results capture humans in a rough shape, with limited elevation resolution.}
    \label{fig:static_rf_imaging_results}
    \vspace{-15pt}
\end{figure}

Fig.~\ref{fig:static_rf_imaging_results} shows a few examples of our 3D imaging outputs using beamforming, together with the reference point clouds captured by a 3D LiDAR. All of the imaging results are generated via one-shot measurement like the LiDAR, i.e., the radar rotates for one cycle. We also visualize a 2D slice (i.e., $z=0$) of the point clouds which resembles 2D mapping results (i.e., floor plans).
It can be observed that our beamforming-based imaging results achieve fine-grained azimuth resolution with a large synthetic aperture and fine-grained range resolution with a large bandwidth at mmWave frequency. However, the elevation resolution remains limited due to a small number of vertical antennas. As a result, beamforming-based 3D images seem over-smoothed and fail to capture the detailed changes along the elevation dimension.

Below we analyze the imaging resolution of our beamforming algorithm, including the first analytical expression for azimuth resolution in circular/cylindrical arrays.

\header{Elevation Resolution.} 
Our design has a linear array along the vertical axis, consisting of $A$ antennas with a spacing of ${\lambda}/{2}$. This configuration results in an elevation resolution of (determined by the 3-dB beamwidth of the imaging mainlobe~\cite{radar_fundamentals}):
$\Delta\theta = 0.89\lambda/{L} = {1.98}/{A}$, where L is the aperture size and the unit of $\Delta\theta$ is in radians.
For the specific radar we use with $A=8$ and $\lambda$ = 3.8~mm, our $\Delta\theta$ is 0.25~rad or 14.2$^\circ$.

\header{Range Resolution.}
The range resolution of FMCM radar is determined by its bandwidth $B$~\cite{radar_fundamentals} as: $\Delta R = {c}/{2B}$, where $c$ is the speed of light. For our radar configuration with $B$ = 4~GHz, our $\Delta R$ is 3.75~cm. 

\header{Azimuth Resolution.}
Our system effectively employs a circular array for resolving reflections along the azimuth dimension. We provide analytic expression for the beam shape and angular resolution when imaging with a circular array. To the best of our knowledge, this is the first set of analytic results for circular arrays. The proof of this lemma follows a similar approach to that of linear arrays~\cite{radar_fundamentals}, with the detailed derivation deferred to Appendix.
\vspace{-3pt}
\begin{lemma}
Consider a circular array of radius $r$ and a reflector at angle $\theta=0$, due to resolution, this reflector influences the imaging of nearby angle $\theta_s$, with voltage $E(\theta_s)$ as:
\begin{equation}
    E(\theta_s) \propto 2\pi J_0 \left(
        \frac{4\pi r}{\lambda} \sin{\frac{\theta_s}{2}} 
    \right),
\label{eq:beamshape}
\end{equation}
where $J_0$ denotes the Bessel function of the first kind~\cite{bessel}. The 3-dB beamwidth of this beam is, therefore, the angular resolution of the circular array:
\begin{equation}
    \Delta\theta = 0.36\frac{\lambda}{r} = 0.72\frac{\lambda}{d},
\label{eq:angular_resolution}
\end{equation}
where $d=2r$ is the diameter of the circular array.
\label{lemma:angular_resolution}
\end{lemma}

With a rotation radius of $r=8$~cm and a wavelength $\lambda=3.8$~mm, \name{} has an azimuth resolution $\Delta\theta = 0.96^\circ$.
We note that Lemma~\ref{lemma:angular_resolution} assumes an omni-directional radiation pattern over either 180 or 360 degrees, the same as the analysis for linear arrays~\cite{radar_fundamentals}. Measurements of the azimuth resolution with actual radar radiation pattern, via both simulations and experiments, are presented in ~\secref{sec:evaluation}.

%% file: src/motion-estimation-5.tex
\section{Motion Estimation and Imaging}
\label{sec:motion_estimation_and_imaging}

\begin{figure}[tbp]
    \centering
    \includegraphics[width=\linewidth]{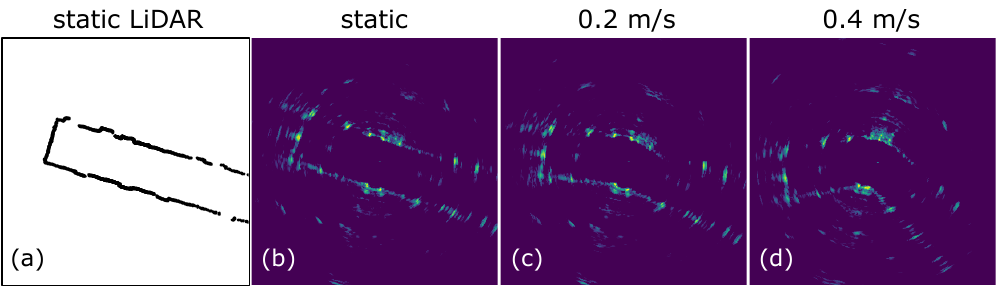}
    \caption{Distortion of the imaging results due to robot motion. 2D visualizations show that the distortion gets worse as the robot starts to move.}
    \label{fig:without_compen}
    \vspace{-15pt}
\end{figure}

The external motion of the platform (e.g., a robot) can introduce unknown motion to the radar in addition to its own rotation. Obtaining antenna locations with sub-wavelength accuracy (i.e., $\lambda/2=1.9$~mm) is crucial for coherent combination and effective beamforming. \figref{fig:without_compen} shows how beamforming (\eqnref{eq:static_imaging}) fails when external motion is not properly estimated and compensated. \figref{fig:without_compen}(a)(b) shows that the 2D images captured by LiDAR and radar are similar when the robot is static. However, when the robot begins to accelerate and move forward (\figref{fig:without_compen}(c)(d)), the RF images become distorted, with objects appearing at incorrect locations.

To address this issue, we leverage the Doppler effect in the reflected signals to estimate robot motion. This task is, however, not trivial, as both the radar rotation and robot motion induce spatial frequency across antennas, leading to a mixed effect due to AoA and Doppler. Below, we describe how we decouple AoA and Doppler effect (\secref{subsec:aoa_dopper}), estimate robot motion (\secref{sec:motion_estimation}), and compensate for it efficiently (\secref{subsec:efficient_imaging}).

\subsection{AoA and Doppler Effect}
\label{subsec:aoa_dopper}
We first model the net motion an antenna undergoes and how this affects its distance to a reflector (the reflectors referred to in this section are objects that already exist in the environment). Assume the antenna rotates with radius $r$ and angular velocity $\omega$ along the z-axis of the robot, and has a linear velocity $\vec{v}$ in the x,y-plane.
We assume both $\omega$ and $\vec{v}$ remain constant over one rotation cycle (i.e., 0.5s).
The origin of the coordinate system is set at the rotation center when $t=0$. We consider a reflector on the x,y-plane with the range $R_n$ and azimuth $ \theta_n$.
The distance between this reflector and the antenna at time $t$ can be approximated (when $R_n$ is much larger than $r$ and $vt$, see Appendix) as:
\begin{equation}
    d(t) = R_n
    - \underbrace{r\cos(\omega t - \theta_n)}_\text{radar rotation}
    - \underbrace{vt\cos(\theta_v-\theta_n)}_\text{robot motion},
\label{eq:dist_scatter_ante}
\end{equation}
where $v$ and $\theta_v$ are the speed and angle of $\vec{v}$.
Notice that the second term in this equation accounts for radar rotation, while the third term results from the robot motion.

\begin{figure}[t]
    \centering
    \begin{subfigure}[b]{0.49\linewidth}
        \includegraphics[width=\linewidth]{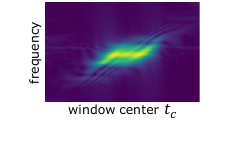}
        \vspace{-35pt}
        \caption{original spectrogram}
    \end{subfigure}
    \hfill
    \begin{subfigure}[b]{0.49\linewidth}
        \includegraphics[width=\linewidth]{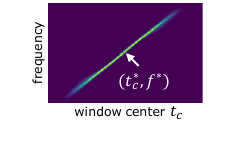}
        \vspace{-35pt}
        \caption{compensated spectrogram}
    \end{subfigure}
    \caption{RF spectrograms showing reflections from a single object. Each column of the spectrogram represents the slow-time FFT within a window centered at a specific $t_c$. The line in (b) follows \eqnref{eq:frequency_ft}.}
    \label{fig:rotation_compen}
    \vspace{-10pt}
\end{figure}

Performing FFT on antenna measurements over time (i.e., slow-time FFT) essentially analyzes the rate at which $d(t)$ changes.
As indicated in \eqnref{eq:dist_scatter_ante}, the slow-time FFT captures a combined effect of AoA and Doppler. Additionally, since $d(t)$ does not vary at a constant rate (i.e., it isn't a linear function of $t$), the slow-time FFT has its energy spread out across spectrum. \figref{fig:rotation_compen}(a) shows the spectrogram obtained by applying FFT over a sliding window whose center is $t_c$.
The primary cause of the dispersion is the cosine nonlinearity introduced by the radar's rotation.
To address this, our approach involves adding a compensation term to $d(t)$ to linearize it in relation to $t$. Specifically, for each sliding window centered around $t_c$, we compensate for $d(t)$ as follows:
\begin{equation}
\begin{aligned}
    d'(t, t_c) &= d(t) + r \cos(\omega(t_c-t)).
\end{aligned}
\label{eq:linear_d}
\end{equation}
This compensation is achieved by multiplying the antenna measurements by $\exp\{j4\pi r\cos(\omega(t_c-t))/\lambda\}$.
For ease of analysis, the compensated distance function $d'(t, t_c)$ can then be approximated (when $\omega t_c-\theta_n$ is relatively small,
see Appendix) as:
\begin{equation}
\begin{aligned}
    d'(t, t_c) = r\omega t(\omega t_c-\theta_n) - vt\cos(\theta_v-\theta_n) + \mathrm{const.},
\end{aligned}
\label{eq:linear_d_final}
\end{equation}
which is linear with respect to $t$. Thus, taking slow-time FFT over compensated antenna measurements would yield a strong response at frequency:
\begin{equation}
    f(t_c) = \frac{2}{\lambda} [r\omega (\underbrace{\omega t_c - \theta_n}_{\mathrm{AoA}}) - \underbrace{v \cos(\theta_v - \theta_n)}_{\mathrm{Doppler\ speed}}],
\label{eq:frequency_ft}
\end{equation}
where the first term reflects AoA (i.e., difference between radar direction and reflector direction) and the second term is the Doppler speed. \figref{fig:rotation_compen}(b) shows the spectrogram of the compensated signals.
Each column is the slow-time FFT result within a window centered at $t_c$, which exhibit sharp response at frequency $f(t_c)$.
Interestingly, it also shows that the $f(t_c)$ is changing linearly over time, as dictated by \eqnref{eq:frequency_ft}.
Consequently, a line persists during the time window when the reflector is within the FoV of the rotating antenna.
Due to the standard antenna radiation pattern~\cite{awr1843beamwidth}, the strongest response appears when the antenna is directly facing the reflector. Specifically, $\omega t_c^* = \theta_n$ if $(t_c^*, f^*)$ denotes the location of the strongest response on this line.
Plugging it into \eqnref{eq:frequency_ft}, we get the following equation regarding peak location:
\begin{equation}
f^* = f(t_c^*) = -2v\cos(\theta_v - \omega t_c^*)/\lambda.
\label{eq:peak_relation}
\end{equation}
Observe that by identifying this line and its peak, we can determine the AoA and Doppler effect separately. Specifically, $\omega t_c^*$ indicates the direction of the reflector, while $-\lambda f^* / 2$ quantifies its Doppler speed.
We use Hough transform~\cite{hough} for line detection in the compensated spectrogram. Since we know the slope for lines of interest (i.e., $2\omega^2r/\lambda$), we speed up the detection by restricting the slope search to this value.

\subsection{Robust Motion Estimation}
\label{sec:motion_estimation}
\begin{figure}[t]
    \centering
    \begin{subfigure}[b]{\linewidth}
        \includegraphics[width=\linewidth]{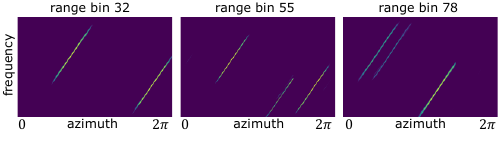}
        \vspace{-25pt}
        \vspace{3pt}
        \caption{Compensated spectrograms at different range bins.}
    \end{subfigure}
    \hfill
    \begin{subfigure}[b]{\linewidth}
        \includegraphics[width=\linewidth]{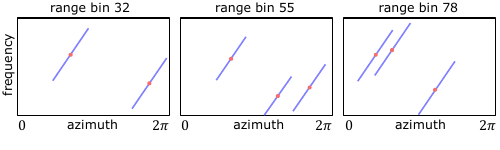}
        \vspace{-25pt}
        \caption{Detected lines and peaks from the compensated spectrograms above.}
        \vspace{3pt}
    \end{subfigure}
    \hfill
        \begin{subfigure}[b]{\linewidth}
        \includegraphics[width=\linewidth]{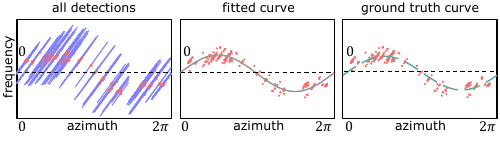}
        \vspace{-25pt}
        \caption{Sinusoidal curve fitting to estimate the speed and heading direction.}
    \end{subfigure}
    \caption{Robust motion estimation with multiple observations. Each line corresponds to a reflector observed within a certain azimuth window.}
    \label{fig:motion_esti}
    \vspace{-15pt}
\end{figure}

Measurement of AoA and Doppler speed from just a single reflector falls short in estimating both $v$ and $\theta_v$.
The Doppler speed, after all, only reveals the radial speed in the direction of the reflector. This is further complicated by the noise or errors in the detected lines and peaks. 
We tackle this challenge by using multiple reflectors together with a robust estimation scheme.
First, we use range FFT to isolate reflectors at different ranges (\figref{fig:motion_esti}(a)).
Multiple reflectors and their corresponding lines are detected (\figref{fig:motion_esti}(b)). 
Notice from \eqnref{eq:peak_relation} that all the peak locations falls on the sinusoidal curve $f = -2v\cos(\theta_v - \omega t_c)/\lambda$, with amplitude propositional to the speed $v$, and initial phase being $\theta_v$.
Therefore, we aggregate all the detected peaks from different azimuth angles and perform a sinusoidal curve fitting to estimate $v$ and $\theta_v$ (\figref{fig:motion_esti}(c)).
We use the RANSAC~\cite{ransac} to leverage the redundancy in our observations and to mitigate the impact of noise and outliers.
It achieves an average error of 8.48~mm/s and 1.09$^\circ$ for $v$ and $\theta_v$ estimations, respectively (\figref{fig:v_esti_error}).

\subsection{Efficient Compensation and Imaging}
\label{subsec:efficient_imaging}
Robot motion introduces an additional displacement $\vec{v}\cdot t$ to the antenna location $\vec{p}^a_t$.
Thus, the beamforming algorithm in \eqnref{eq:static_imaging} can be revised into:
\begin{equation}
    \boldsymbol{B}(\vec{d}, \vec{v}) = \sum_{a,t} \boldsymbol{S}^a_t 
        \exp \left( j2\pi \frac{2\vec{d} \cdot (\vec{p}^a_t + \vec{v}\cdot t)}{\lambda} \right).
\label{eq:motion_imaging}
\end{equation}
\figref{fig:moving_rf_imaging_results} shows our imaging results with a moving robot, where the environment is captured accurately without distortion.
We analyze the complexity of different algorithms below. $\Theta$ and $\Phi$ denote the number of azimuth and elevation angles for imaging, $W$ represents the number of antennas within the sliding window, $A$ is the number of antennas vertically, and $N$ indicates the size of FMCW chirp.

\header{Phase Steering.}
Using Delay-and-sum~\cite{radar_fundamentals, DAS_beamforming} in our system would have a computational complexity of $O(\Theta \Phi W A N^2)$.

\header{2D Beamforming.}
Using beamforming \eqref{eq:motion_imaging} followed by range FFT has a complexity of $O(\Theta \Phi W A N \log N)$.

\header{Consecutive 1D Beamforming.}
We re-write \eqref{eq:motion_imaging} into:
\begin{equation}
\begin{aligned}
    \boldsymbol{B}(\vec{d}, \vec{v}) &= 
    \sum_t \boldsymbol{S}_t' \exp \left(j4\pi \vec{d}' \cdot (\vec{p}_t' + \vec{v}' \cdot t) / \lambda \right),\\
    \boldsymbol{S}_t' &= \sum_a \boldsymbol{S}^a_t \exp \left(j4\pi h^a \sin\phi_d / \lambda \right),
\end{aligned}
\label{eq:motion_imaging_efficient}
\end{equation}
where $\vec{d}'$, $\vec{p}_t'$ and $\vec{v}'$ are the first two dimensions of $\vec{d}$, $\vec{p}^a_t$ and $\vec{v}$, respectively.
It shows that the compensation in elevation is independent of that in azimuth.
This approach is effectively performing two consecutive steps of 1D beamforming, with a total complexity of $O(\Theta \Phi A N + \Theta \Phi W N\log N) = O(\Theta \Phi W N\log N)$ given that $W \gg A$.

%% file: src/3d-ML-4.tex
\section{Enhanced Imaging with ML}

\begin{figure}[t]
    \centering
    \includegraphics[width=\linewidth]{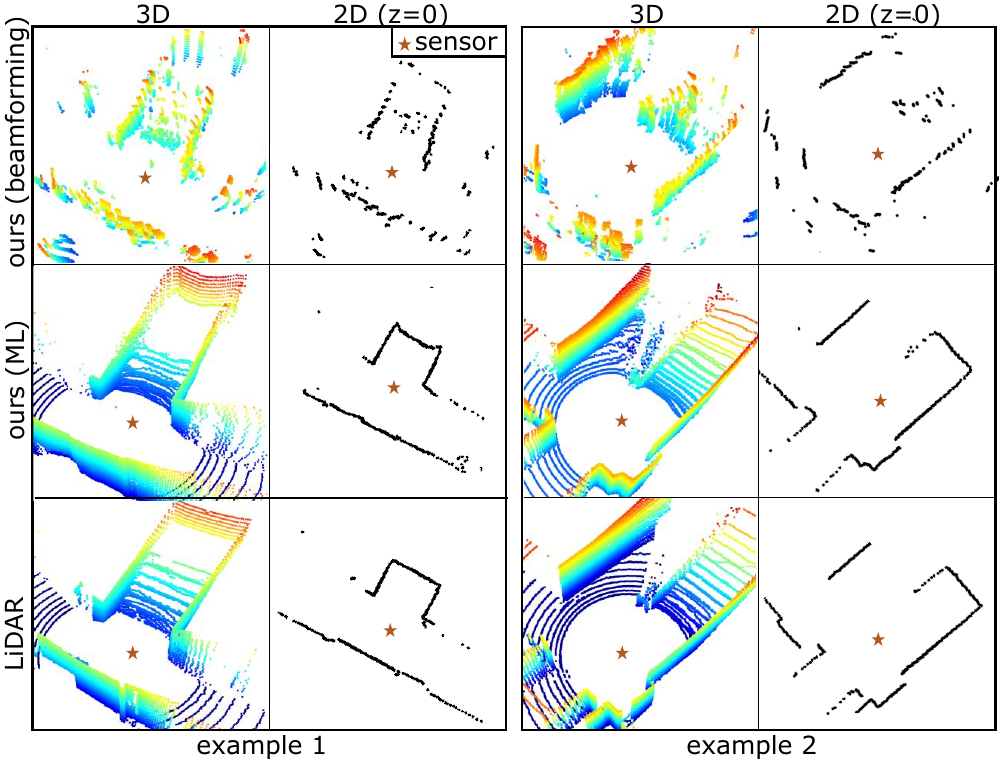}
    \caption{RF imaging results with a moving robot with beamforming (top) and with additional ML-based resolution enhancement (mid). Our motion estimation and compensation avoid the distortion, resulting in high range and azimuth resolutions. The 3D learning model further enhances the elevation resolution, showing detailed structures like stairs.}
    \label{fig:moving_rf_imaging_results}
    \vspace{-10pt}
\end{figure}

Through signal processing alone, \name{} has achieved fine-grained resolution in both the azimuth and range dimensions. However, the elevation resolution remains limited, especially when compared to the other dimensions (\figref{fig:moving_rf_imaging_results}).
While this difference seems challenging to mitigate, it offers a unique opportunity when viewed through the lens of machine learning.
Given the structural properties inherent in 3D environments, spatial dimensions are not independent. For instance, consistent depth cues from surfaces, as well as gravity constraints (e.g., humans and objects need support and stand on the floor), provide cross-dimensional information. A CNN can utilize these cross-dimensional dependencies to resolve the reflectors on the poor elevation dimension. For instance, with stairs, each step's distinct distance from the radar naturally separates it from other steps. 
In this section, we describe the design of our model to enhance the elevation resolution and address the challenges arising from the unique characteristics of RF signals (\secref{sec:range_estimation_ml}). 
We also describe various downstream applications enabled by our high-resolution RF images (\secref{sec:visual_recog_model}), as well as panoramic learning (\secref{sec:panoramic_learning}) and our methodology to collect visual recognition labels (\secref{sec:ground_truth_labels}) of our dataset that we will release to public for broader impact.

\subsection{Resolution Enhancement with ML}
\label{sec:range_estimation_ml}
\begin{figure}[t]
    \centering
    \includegraphics[width=\linewidth]{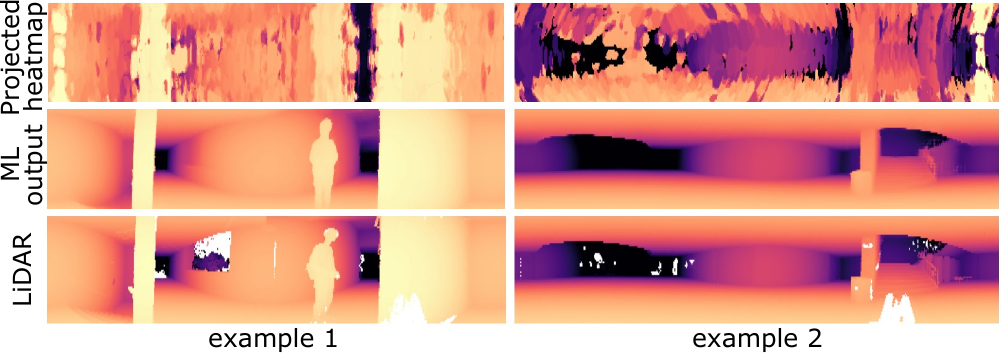}
    \vspace{-12pt}
    \caption{3D RF heatmap (top) and the resolution enhancement model output (middle).
    To visualize a 3D heatmap, we take the range value of the peak in each direction and visualize the 3D heatmap as a 2D image.}
    \label{fig:network_input}
    \vspace{-8pt}
\end{figure}

Our learning adopts a cross-modal strategy, with paired RF and LiDAR data as inputs and targets for the training. 
We compute RF inputs using beamforming (\secref{sec:motion_estimation}).
We query beams to match the LiDAR's grid of azimuth and elevation angles.
This results in RF tensors of size 512 x 64 x 256 (i.e., azimuth x elevation x range). 
It's important to note that the number of elevation angles doesn't reflect the elevation resolution, as evidenced by the over-smoothing observed along the elevation dimension (\figref{fig:network_input}, top).
Our learning target from 3D LiDAR can be viewed either as sparse 3D point clouds or as dense 2D range maps of size 512 x 64.

\header{3D Learning via 2D Convolutions.} Unlike previous studies~\cite{radatron, CMU_mmWave_pc, RF-pose} that treat range as a spatial dimension for convolution, our model regards the range as the channel dimension. This allows us to process our 3D RF inputs with 2D CNN models. While conventional CNNs typically expand channel dimensions in the initial layer to extract high-dimensional features, our model aims to compress the sparse signals along the range dimension (by 4x), enhancing learning efficiency. Furthermore, our model adopts the 2D range map representation of LiDAR data as the learning target (\figref{fig:network_input}, bottom), enabling the cross-modal learning to operate with 2D convolutions.
Compared to its 3D CNN counterpart, our 2D model is more memory and computationally efficient. Moreover, 2D CNNs are less prone to overfitting when learning sparse 3D structures from sparse 3D RF inputs.

\begin{figure}[t]
    \centering
    \begin{minipage}[t]{1.65in}
        \includegraphics[width=\linewidth]{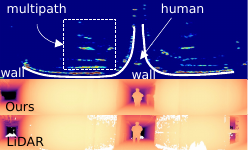}
        \caption{\name{} learns the first reflection under the presence of multipath interference.}
        \label{fig:multi_path}
    \end{minipage}
    \hfill 
    \begin{minipage}[t]{1.65in}
        \includegraphics[width=\linewidth]{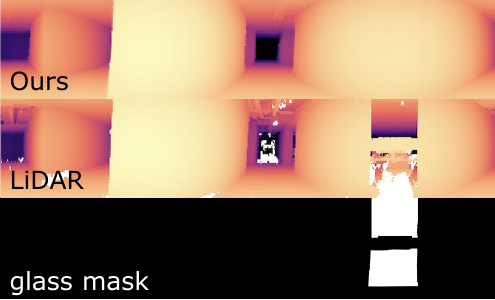}
        \caption{\name{} properly handles glass regions (bottom, white pixels) and recovers their depth (top) while LiDAR fails to do so (middle).}
        \label{fig:glass}
    \end{minipage}
    \vspace{-8pt}
\end{figure}

\begin{figure}[t]
    \centering
    \includegraphics{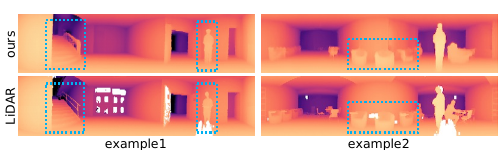}
    \vspace{-12pt}
    \caption{\name{} produces range images with fine-grained details. }
    \label{fig:details}
    \vspace{-8pt}
\end{figure}

\header{Handling Multipath Effect.}
Learning the range of the first reflector as captured by LiDAR yields an opportunity to mitigate the multipath effect that is predominant in indoor environments.
By emulating a modality less susceptible to multipath effects, our RF-based model learns to associate the reflections in the RF data to be able to mitigate multipath interference.
\figref{fig:multi_path} shows a multipath-rich corridor where our system accurately senses the 3D environment and  the person, without being affected by the multipath reflections.

\header{Handling Glass.}
During cross-modality learning, it is crucial to account for the intrinsic differences between how RF and LiDAR interact with glass (e.g., windows, glass doors).
While LiDAR cannot detect glass due to its optical properties, glass is opaque to mmWave signals~\cite{glass_mmWave}.
We incorporate glass masking and employ masked L1 loss to ensure our RF-based model remains robust and isn't misled by supervisory signals in those regions.
More specifically, the loss computations within the masked glass regions will be ignored, so that the model will not be impacted by the incorrect supervision from the LiDAR sensor for these regions.
The glass masks were collected together with other semantic labels for segmentation described in \secref{sec:ground_truth_labels}.
As a result of its training, \name{} successfully recovers the depth of transparent glass (\figref{fig:glass}), showcasing the potential of RF imaging to enhance robotic collision avoidance by not being fooled by transparent obstacles like glass doors or windows.

\header{Capturing Details.}
L1 loss tends to produce median-like effects~\cite{l1_median}, which can lead to smooth predictions.
Training with L1 loss alone would fail to capture high-frequency details.
We incorporate perceptual loss into our training.
This ensures a more faithful recovery of high-fidelity range images, which is crucial for effective visual recognition tasks. 
We employed LPIPS~\cite{lpips} with learnable weights for features at different layers.
\figref{fig:details} shows that our approach effectively captures object details including human and stairs.

\vspace{-13pt}
\subsection{Visual Recognition with ML}
\label{sec:visual_recog_model}
With enhanced RF resolution close to that of LiDAR, \name{} captures a comprehensive understanding of the surroundings. This enables various downstream tasks, including the first set of visual recognition applications with RF signals.

\header{Surface Normal} is crucial for visual perception and robotics tasks like SLAM~\cite{vi-slam}. Obtaining an accurate range map is essential for capturing surface normals, as small errors in depth can result in significant deviations in the normal directions.
We add an additional convolutional layer to our resolution-enhancement model to predict surface normal vectors.
We use LiDAR to derive ground truth for training~\cite{surface_normal}.
To visualize surface normal, we use a standard color mapping that linearly maps the xyz values to RGB~\cite{cvbook}.

\header{Semantic Segmentation} provides pixel-level scene understanding. 
We employ one of the state-of-the-art methods~\cite{DeepLabv3+} that does not involve transformers. We observed that vision transformers did not provide additional benefits, possibly due to the small elevation dimension (i.e., 64) of our data. We use a pre-trained ResNet-101 as our backbone~\cite{ResNet}, which transfers the rich multi-scale features present in natural images to our RF-based range images. Our model is trained to predict 11 semantic classes: table/chair, human, trashcan, railing, door, elevator, stairs, wall, window, floor, and ceiling.

\header{Object Detection} has applications in robot navigation, warehouse management, HCI, etc. 
We use the same ResNet backbone for semantic segmentation, with a Feature Pyramid Network (FPN)~\cite{Feature-Pyramid-Networks} and a Faster R-CNN~\cite{faster-rcnn} for detection.
Unlike \cite{see_through_smoke} that has constraints on the radar's position and classifies objects given a predefined region, we treat object detection as a general task on the RF imaging results.
Our objective is to accurately predict bounding boxes for human class and other non-human objects within the scene.

\header{Human Localization} could be achieved as a byproduct of object detection. Specifically, we infer azimuth from human bounding box coordinates, and range from the predicted depth inside this bounding box (\figref{fig:human_loc}). This approach provides a novel perspective of solving RF-based indoor human localization, as it inherently overcomes the multi-path problem that 
is challenging in device-free localization~\cite{mdtrack, rssi, fingerprinting}. 

\subsection{Panoramic Learning}
\label{sec:panoramic_learning}

A unique feature of our input images is their inherent panora-mic nature. Our model therefore has the opportunity to be optimized to utilize features that cross the left and right boundaries of the input. Traditional learning methods will not be able to identify and take advantage of these features. For example, an object could be split at the boundary of an image, but current detection models would predict two separate bounding boxes at each end with possible misclassification, or miss completely since neither split provides enough features for recognition (the person in \figref{fig:pano_det}). This will lead to a decreased detection performance~\cite{panoramic-detection}.

We make the following changes to our model to take advantage of the panoramic nature of our data:
1) apply circular padding instead of typical zero padding along azimuth for all convolution layers;
2) disable bounding box clamping for azimuth dimension;
3) revise IoU calculation by taking cross-boundary bounding boxes into account; 
4) modify Region of Interest (ROI) pooling technique to first duplicate the feature maps horizontally before applying the pooling. We evaluate the impact of panoramic learning and find it provides consistent gains for all our learning tasks (\secref{sec:evaluation}).

\subsection{Ground Truth Labels}
\label{sec:ground_truth_labels}
For semantic and object labels, we use an active learning~\cite{al} strategy. We started with Segment Anything Model~\cite{SAM}, a promptable tool, to aid the manual annotation of semantic labels. Meanwhile, we extracted object boxes by taking the semantic connected components with manual correction. This process was applied to the first 10\% of the data.
Next, we developed an semantic and object annotation model identical to the one in \secref{sec:visual_recog_model} to generate semi-annotated data, which again underwent human corrections. We refined the model iteratively with additional corrected annotations. With each iteration, the model's accuracy improved, diminishing the need for human correction. Upon reaching 50\% annotations, we deployed the model to infer labels on the remaining data. 
Collectively we have 11,033 semantic and surface normal annotated images, together with 20,546 object instances.

%% file: src/implementation.tex
\begin{table*}[h]
    \footnotesize
    \caption{The building information (sorted by year of construction) and our system's cross-building generalization performance.}
    \centering
    \begin{tabular}{c|c|c|c|c|c|c|c|c|c|c|c|c}
        \hline
        Building \# & 1 & 2 & 3 & 4 & 5 & 6 & 7 & 8 & 9 & 10 & 11 & 12 \\
        \hline
        Year of Construction & 1906 & 1913 & 1925 & 1940 & 1954 & 1966 & 1967 & 1971 & 1987 & 1996 & 2006 & 2013  \\
        Year of Interior Renovation & 2017 & 2017& 2005 & 1973 & 1967 & 2003 & 2013 & 1999 & N/A & N/A & N/A & N/A \\
        \hline
        
        Range Image MAE (cm) & 12.62 & 13.77 & 16.40 & 20.25  & 11.19 & 13.34 & 10.36 & 30.27 & 12.23 & 16.24 & 16.12 & 19.20 \\
        Surface Normal MAE ($^\circ$) & 7.63  & 8.68  & 9.65 & 10.17  & 6.88 & 8.83 & 9.47  & 9.92 & 9.31  & 9.90 & 8.27 & 8.79 \\
        Object Detection $\mathrm{AP^{30}}$ & 57.20  & 52.11  & 64.88 & 35.64  &  51.17 & 67.47 & 65.48 & 44.20 & 58.50 & 61.71 & 65.97 & 50.03 \\
        Semantic Segmentation mIoU & 51.11 & 51.03 & 50.07 & 47.02  & 52.23 & 46.30 & 38.77 & 45.69 & 53.58 & 45.64  & 39.50 & 38.87 \\
        \hline
    \end{tabular}
    \label{tab:lobo}
\end{table*}

\begin{table*}[h]
    \footnotesize
    \caption{The overall performance of our surface normal estimation, object detection, semantic segmentation and 2D human localization.}
    \centering
    \begin{tabular}{ccccccccccccc}
    \hline
     \multicolumn{3}{c}{Surface Normal Error} & & \multicolumn{2}{c}{Object Detection} & & \multicolumn{2}{c}{Semantic Segmentation} & & \multicolumn{2}{c}{Human Localization Error} \\
    \cline{1-3}
    \cline{5-6}
    \cline{8-9}
    \cline{11-12}
     mean & median & 90th-percentile & & $\mathrm{AP^{30}}$ & $\mathrm{AP^{50}}$ & & mIoU & pAcc &  & mean & median \\
    8.83$^\circ$ & 2.17$^\circ$ & 28.10$^\circ$ & & 52.34 & 38.30 &  & 48.00 & 86.33 & & 12.24 cm \& 1.47$^\circ$ & 5.78 cm \& 1.08$^\circ$ \\
    \hline
    \end{tabular}
    \label{tab:vr}
    \vspace{-5pt}
\end{table*}

\section{Implementation and Dataset}
\label{8. Implementation and Dataset}
In this section, we describe the implementation details of \name{} and the dataset used for training and evaluation.

\begin{figure}[t]
    \centering
    \includegraphics[width=0.7\linewidth]{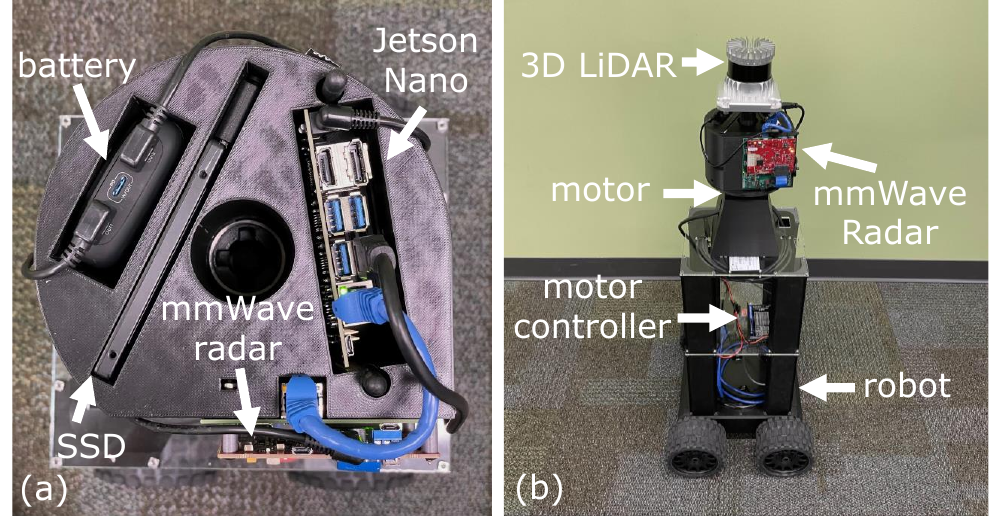}
    \caption{Implementation. Left: hardware setup that captures RF signals. Right: \name{} and LiDAR mounted on a mobile robot platform.}
    \label{fig:system_img}
    \vspace{-13pt}
\end{figure}

\header{Hardware.} \figref{fig:system_img}(a) shows the hardware design of our system. We use a TI AWR1843 single-chip mmWave FMCW radar~\cite{awr1843beamwidth}, with a DCA1000EVM data capture board~\cite{DCA1000}.
We configure our radar to sweep from 77 to 81 GHz with a bandwith of 4 GHz. Each chirp has 256 samples, with a 10~m maximum sensing range.
A single board computer (i.e., Jetson Nano) records the raw samples.
We use a Nema 23 stepper motor to drive the rotary part at 2Hz.
The rotation radius is set at 8cm.
Besides, an Ouster 64-beam LiDAR is used to provide ground truth.
All cases and supporting parts are 3D printed.
Our system is mounted on a Lynxmotion Mecanum Rover mobile platform, as shown in Fig.~\ref{fig:system_img}(b).

\begin{figure}[t]
\label{panofig}
    \centering
    \begin{minipage}[t]{1.65in}
        \includegraphics[width=\linewidth]{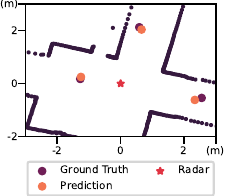}
        \caption{\name{} localizes humans on a 2D floor plan. }
        \label{fig:human_loc}
    \end{minipage}
    \hfill 
    \begin{minipage}[t]{1.65in}
        \includegraphics[width=\linewidth]{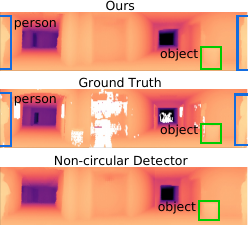}
        \caption{\name{} detects objects across boundary.}
        \label{fig:pano_det}
    \end{minipage}
\vspace{-10pt}
\end{figure}

\header{Neural Networks.}
Our resolution enhancement model is structured into 7 stages, each with 4 ResNet blocks. The channel count for each stage is determined by a factor of (1,2,4,8,8,8,8) relative to the stem output channels.
We train our model using AdamW with an initial learning rate of $10^{-3}$, decaying by a factor of 0.1 at 50k and 80k iterations. Weight $\alpha$ for perceptual loss is 0.1.
Our semantic segmentation model is trained using hard-pixel-mining loss~\cite{deeplab} with label smoothing of 0.1.
For object detection, we use a NMS threshold of 0.7 during training, and 0.5 during inference.

\begin{figure}[t]
    \centering
    \includegraphics[width=0.90\linewidth]{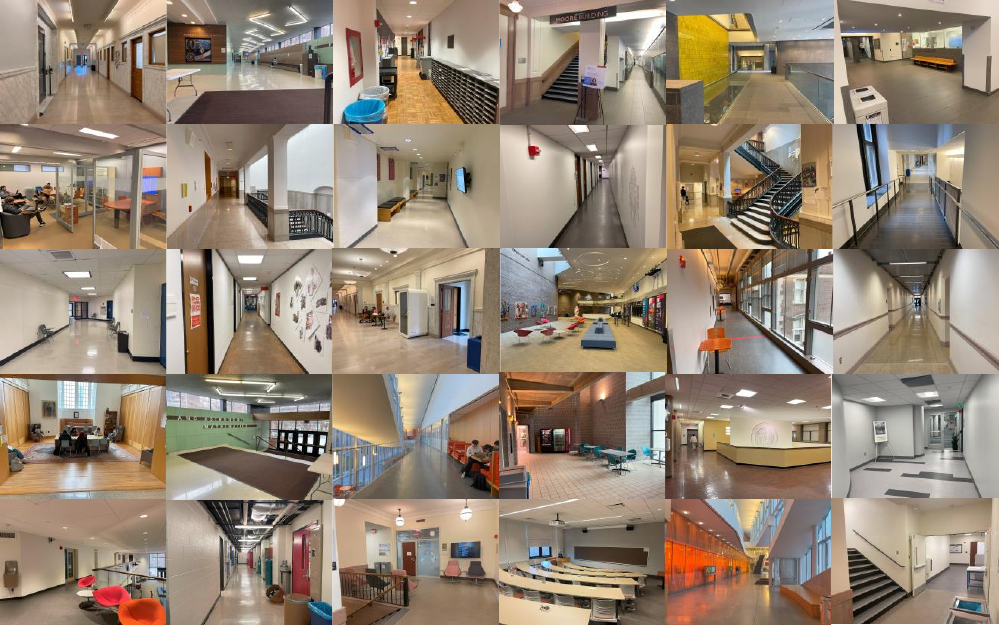}
    \caption{RGB images showing different environments in the dataset.}
    \label{fig:rgb_examples}
    \vspace{-12pt}
\end{figure}

\header{Dataset.} We collected a large dataset that includes a total of 11,033 synchronized RF and LiDAR data spanning across 12 distinct buildings, constructed over a span of a century (from 1906 to 2013). 
Each building possesses unique features, showcasing materials and designs indicative of their respective eras. 
\figref{fig:rgb_examples} shows example scenarios in our dataset.
The dataset after processing amounts to 461 GB.

%% file: src/evaluation-3.tex
\begin{figure*}[h]
    \centering
    \includegraphics[width=\linewidth]{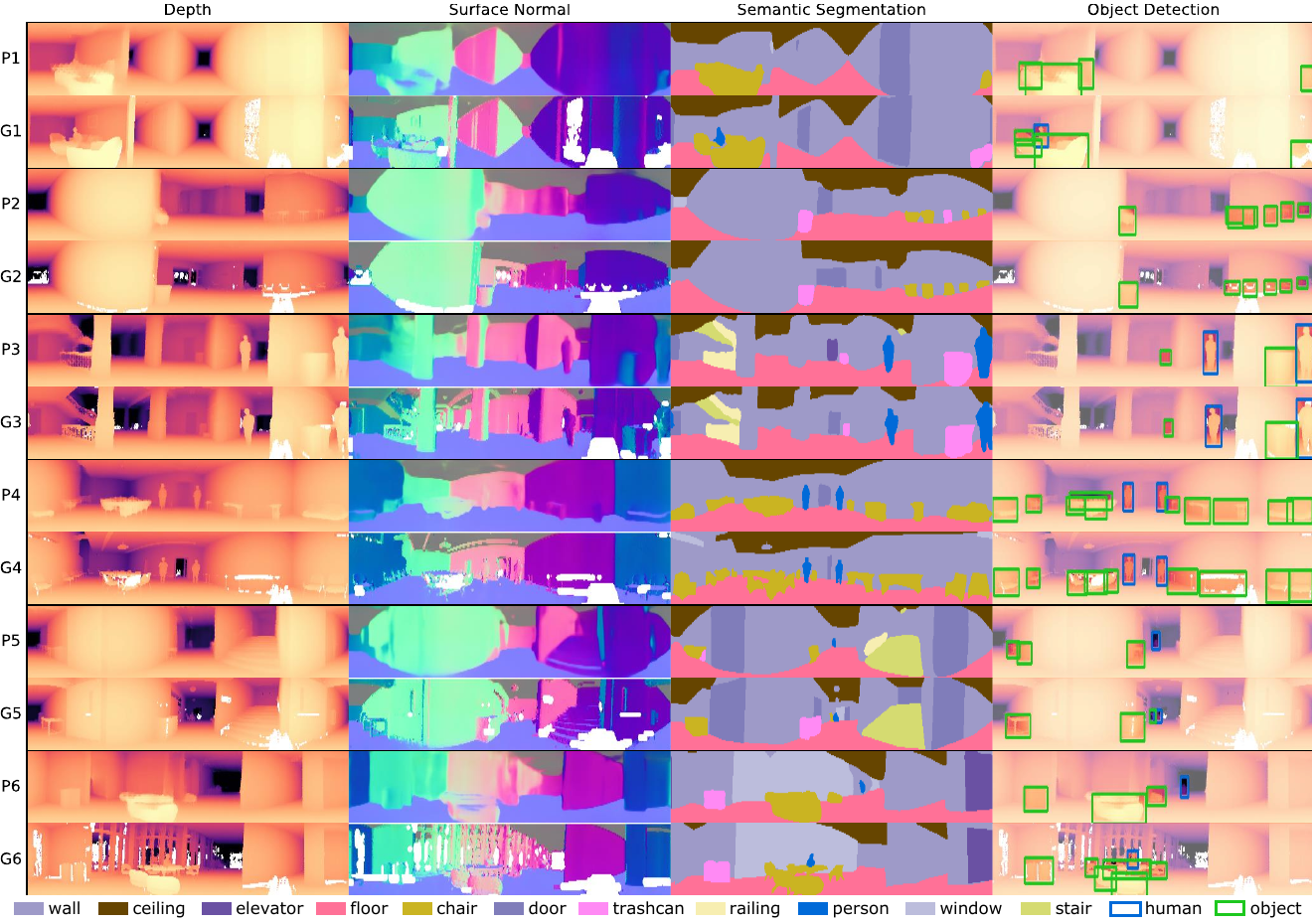}
    \caption{\name{} performance on diverse building environments. For every two rows, the top is our prediction and the bottom is the ground truth.}
    \label{fig:gallery}
    \vspace{-5pt}
\end{figure*}

\section{Evaluation}
\label{sec:evaluation}
In this section, we evaluate the performance of \name{}.

\header{Training \& Testing Split.}
We evaluated all machine learning methods with a cross-building approach to ensure the generalization of our model. Specifically, we left out each building for testing while using the rest for training, repeating this process 12 times, once for each building.

\begin{figure}[htbp]
    \centering
    \begin{minipage}[t]{1.65in}
        \includegraphics{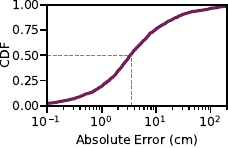}
        \caption{The CDF for absolute error of range image estimation.}
        \label{fig:depth_building}
    \end{minipage}
    \hfill 
    \begin{minipage}[t]{1.65in}
        \includegraphics{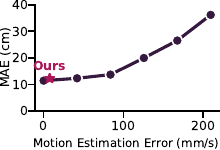}
        \caption{The effect of motion errors to imaging performance.}
        \label{fig:speed_error}
    \end{minipage}
    \vspace{-10pt}
\end{figure}

\header{Range Image Accuracy.}
The range image is the core output of \name{} for capturing the surrounding 3D structure.
When compared against LiDAR ground truth, our model achieves a mean absolute error (MAE) of 15.76 cm.
\figref{fig:depth_building} shows the cumulative distribution function (CDF) of the error for the predicted range image. Notably, the median error is a mere 3.39 cm, suggesting that over half of the predictions are within this tight error margin. Our 90th-percentile error is 31.98 cm, and a large portion of the tail error can be attributed to inaccurate boundaries between foreground and background objects, which has limited impact for applications that do not require pixel-level accuracy. \tabref{tab:lobo} presents per-building performance for this task, showing that our model is robust to variations in the building environments.

\header{Point Cloud Accuracy.}
An alternative method of evaluating our predicted range results is to compare them against the ground truth in a 3D point cloud space. CFAR \cite{radar_fundamentals} is used to retrieve the point clouds from RF heatmaps, and this intermediate result is also evaluated against the ground truth. We evaluate our system using mean Chamfer Distance (CD) \cite{chamfer} and a mean modified Housdorff Distance (HD) \cite{hausdorff}, both of which are commonly used methods for point cloud comparisons. These errors are summarized in \tabref{tab:rf_imging_eval}.
Overall, our system obtains highly accurate 3D point clouds at a mean CD of 6.96 cm and a mean modified HD of 3.23 cm.
When comparing the results with and without machine learning, we see that machine learning results provides significant gains in performance compared to the signal processing results (\tabref{tab:rf_imging_eval}). This verifies that the deep model helps mitigate artifacts such as multipath and side lobe leakage captured in the raw signal, significantly improving the sensing resolution and refines the imaging results.

\begin{table}[t]
    \footnotesize
    \caption{RF imaging point cloud error with and without machine learning. (CD: Chamfer Distance, HD: Modified Housdorff Distance)}
    \begin{tabular}{ccccc}
    \hline
            & CD (2D) & CD (3D) & HD (2D) & HD (3D) \\
    \hline 
        Beamforming only &  21.4 cm  &  26.6 cm  & 12.8 cm  & 12.0 cm    \\
        Beamforming + ML  &  7.43 cm  &  6.96 cm  & 3.12 cm  & 3.23 cm   \\
    \hline
    \end{tabular}
    \label{tab:rf_imging_eval}
    \vspace{-12pt}
\end{table}

\header{Visual Recognition Performance.} We evaluate the performance of visual recognition with the enhanced RF imaging resolution. For surface normal estimation, we observe a MAE of 8.83$^\circ$ and a notably lower median error of 2.17$^\circ$ (\tabref{tab:vr}). 
Such precision allows for advanced indoor robotics applications that require surface analysis, such as 3D scene reconstruction.
In addition, our semantic recognition model achieves a mean Intersection over Union (mIoU) of 48.00 across 11 classes, demonstrating that our RF images capture rich information and enable ML models to recognize distinct semantic regions. It enables advanced RF-based scene understanding where ML models can further analyze surfaces with different characteristics. Finally, our object detector achieves an $\mathrm{AP^{30}}$ score of 52.33, and an $\mathrm{AP^{50}}$ of 38.30, making it suitable for applications like collision avoidance.

\header{Human Localization Performance.} Our system can perform 2D human localization and achieves an average error of
12.24 cm along range and 1.47 degrees along azimuth~(\tabref{tab:vr}). This performance is comparable to state-of-the-art device-free localization methods~\cite{widar2, mdtrack}. This level of precision can be applied to various applications like patient monitoring and smart home automation.

\header{Qualitative Results.} \figref{fig:gallery} shows qualitative results of our system's performance across many diverse buildings. We remark that the RF-based range images preserve high-frequency details (e.g., chair, railing, stairs) without many of the artifacts of LiDAR (e.g., failed regions, cannot handle transparent surfaces, cannot see through smoke, etc). For example, the details on the stairs in row P5 are well defined, which translates to impressive surface normal and segmentation results. Furthermore, small objects like the chairs in P2 are also seen by our range image, leading to accurate object detection. Our model also demonstrates robustness to dense environments, as evident in P4, where many chairs close together are segmented and detected with high precision.

\begin{figure}[t]
    \centering
    \includegraphics{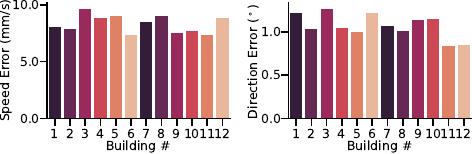}
    \caption{Error of motion estimation across 12 buildings.}
    \label{fig:v_esti_error}
    \vspace{-10pt}
\end{figure}

\header{Motion Estimation Accuracy.}
We evaluate our proposed robot motion estimation by comparing it with the LiDAR ground truth derived using iterative closest point~\cite{ICP}.
Our robot operates with a maximum speed of 0.6 m/s and an average speed of 0.39 m/s, typical for indoor robots.
Fig. \ref{fig:v_esti_error} shows the evaluation results for the 12 individual buildings.
The MAE of speed estimation is 8.48 mm/s and that of the direction estimation is 1.09$^\circ$.
Our motion estimation errors, both in terms of the speed and direction, remain similar across these diverse environments.
This shows that our method is not only accurate (achieving millimeter-level accuracy), but also very robust, a key feature for robotic applications.

\header{Imaging Performance versus Motion Estimation Errors.}
To understand how our imaging model performs with different amounts of error in motion estimation, we introduce synthesized motion estimation errors incrementally to the ground truth and evaluate the model's performance. Fig.~\ref{fig:speed_error} shows that it is resistant to certain amount of motion estimation error. Given the actual performance of our motion estimation algorithm (8.48~mm/s), our ML models are able to maintain robust and accurate across a variety of scenarios.

\begin{figure}[t]
    \centering
    \includegraphics{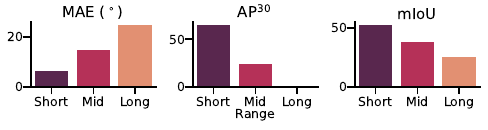}
    \caption{Performance of visual recognition at different distances: short (0-3m), mid (3-6m) and long (6-10m).}
    \label{fig:sub-distance}
    \vspace{-15pt}
\end{figure}

\header{Imaging Performance versus Sensing Distance.}
Based on the distances relative to the radar's position, we split our imaging performance into 3 categories--short, mid and long as depicted in \figref{fig:sub-distance}. A clear trend emerging from the data is that performance drops off with greater distances. This is inevitable as our RF sensing resolution is still affected by the reduced signal strength and low SNR at long range. However, some tasks are more resilient. For instance, semantic segmentation may use semantic features to infer nearby pixels, maintaining continuity. For indoor applications or environments where long-range accuracy is not critical, our system demonstrates its strong performance.

\begin{figure*}[htbp]
    \centering
    \begin{minipage}[t]{5.175in}
        \includegraphics[width=\linewidth]{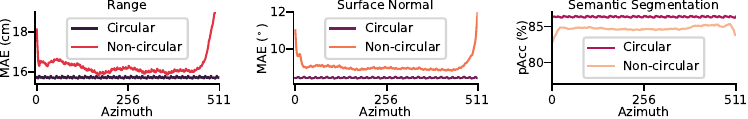}
        \caption{The column-wise metrics of circular and non-circular model across 512 azimuth columns for range estimation, surface normal estimation, and semantic segmentation.}
        \label{fig:column_wise}
    \end{minipage}
    \hfill 
    \begin{minipage}[t]{1.6in}
        \includegraphics[width=\linewidth]{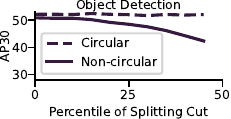}
        \caption{The $\mathrm{AP^{30}}$ verses split percentile in panoramic-rotate test. }
        \label{fig:pano_test}
    \end{minipage}
    \vspace{-10pt}
\end{figure*}

\header{Effect of Panoramic Learning.}
\figref{fig:column_wise} shows the impact of panoramic learning for range imaging, surface normal, and semantic segmentation. To evaluate this, we rotated each test image horizontally 512 times, one column at a time and averaged the errors column-wise. As depicted, our circular model maintains a constant high performance across all orientations of the image, whereas the non-circular model sees drops in performance towards the horizontal edges.

We design a separate test to evaluate our panoramic object detection. Since our dataset does not feature many bounding boxes crossing the boundary, we intentionally rotate our test images to ensure each image presents at least one bounding box that is split at the image boundary. We vary the percentile of box width on which we split the boxes and assess the performance of both circular and non-circular models. As shown in \figref{fig:pano_test}, our results indicate that circular model performs consistently well across various percentiles of partial boxes on both ends, given its ability to always capture the complete circular scene. In contrast, the non-circular model exhibits a decrease in performance, with the most significant drop observed when the split occurs at the 50th percentile. These experiments demonstrate the strengths of utilizing the unique circular features of our input.

\begin{figure}[t]
    \centering
    \includegraphics{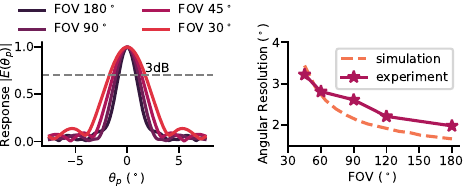}
    \vspace{-3pt}
    \caption{Left: simulated response $E(\theta_s)$ with different FOV window sizes. Right: simulation and experiment results for angular resolution.}
    \vspace{-12pt}
    \label{fig:angular_res}
\end{figure}

\header{Angular Resolution.}
\secref{sec:static} analyzes the angular resolution of circular arrays with omni-directional antennas.
However, the antennas we used in the system are directional and thus have a limited FOV.
To simulate the response $E(\theta_s)$ of the point reflector in \eqnref{eq:angular_resolution}, we adopt the antenna radiation pattern given in~\cite{awr1843beamwidth}.
\figref{fig:angular_res} (left) presents the simulation of the response $E(\theta_s)$ with different FOVs.
With a larger FOV, the beamwidth of the main lobe gets smaller, indicating a better angular resolution.
However, as the FOV reaches angles greater than 90$^\circ$, the rate of shrinkage of the beamwidth decreases.
Since a larger FOV also requires more operations, we make a trade-off between imaging speed and resolution, choosing 90$^\circ$ as the FOV for our system.

To test the azimuth angular resolution of our system, we place two corner reflectors 1 meter away from the radar.
We then record the minimum angle that separates the two individual peaks with its 3-dB beamwidth in the imaging result as the azimuth resolution.
As shown in \figref{fig:angular_res} (right), the angular resolution from experiments is close to that in the simulation.

\begin{table}[t]
    \footnotesize
    \caption{Floating point operations (FLOPs) of our system.}
    \centering
    \begin{tabular}{lllll}
    \hline
    \multicolumn{1}{c}{}  & \multicolumn{2}{c}{Signal Processing} & \multicolumn{2}{c}{ML} \\ 
    \cmidrule(lr){2-3} \cmidrule(lr){4-5}
    \multicolumn{1}{c}{} & \multicolumn{1}{c}{Motion Estimation} & \multicolumn{1}{c}{ Beamforming} & \multicolumn{1}{c}{Range} & \multicolumn{1}{c}{Downstream}\\ \hline
    \multicolumn{1}{c}{GFLOPs} & \multicolumn{1}{c}{0.24} & \multicolumn{1}{c}{11.36} & \multicolumn{1}{c}{103.51} & \multicolumn{1}{c}{165.92} \\ \hline 
    \end{tabular}
    \vspace{-15pt}
\label{tab:runtime}
\end{table}

\header{Runtime Analysis.}
\tabref{tab:runtime} shows the number of floating point operations for various components of our system.
When using desktop GPUs (NVIDIA RTX 3090), it takes 51 ms to perform range estimation, and 44 ms to perform all the downstream visual recognition tasks, totaling 95 ms. The computation takes 726 ms on single board computers (Jetson Orin Nano) and 65 ms on server grade GPUs (NVIDIA L40).

%% file: src/limitations.tex
\section{Limitations and Future Work}
While \name{} has advanced the capabilities of RF imaging systems, it is important to understand its limitations. First, our study has been focused on indoor robots and environments. Although we anticipate that this approach could generalize to other settings, such as warehouses, shopping malls, and even autonomous driving scenarios, these applications remain exciting topics for future studies. Second, the commercial COTS radar used in our system has only eight antennas in an array. Utilizing a radar with a greater number of antennas would lead to higher elevation resolution, potentially enhancing the overall accuracy of the system or achieving the same level of resolution with smaller ML models. 
Lastly, our system is designed to learn first-reflection range images and is trained to ignore multipath reflections. As a result, it cannot leverage these reflections for applications that require seeing through or around capabilities. Developing a method to take advantage of multipath reflections while maintaining the efficiency and robustness of learning presents an interesting avenue for future research.

%% file: src/conclusion.tex
\section{Conclusion}
\name{} introduces a novel RF imaging approach that narrows the resolution gap between RF and LiDAR, enabling a range of visual recognition tasks at radio frequency, including surface normal estimation, semantic segmentation, and object detection. These new capabilities, combined with high-resolution 3D images of the surroundings, open up a multitude of applications that were traditionally only possible with cameras. We believe \name{} marks a significant step forward in RF imaging. We anticipate that this work, along with the released dataset, will encourage further research and development in RF-based imaging technologies, providing a robust yet cost-effective alternative to existing imaging technologies such as LiDAR and cameras.

%% file: src/appendix.tex
\appendix
\section{Derivation of Equations}
\label{sec:appendix_equations}

\header{Equation \ref{eq:beamshape}.}
Similar to the analysis for linear arrays~\cite{radar_fundamentals}:
\begin{equation}
\begin{aligned}
    E(\theta_s) &= \smallint_0^{2\pi} \frac{\exp \left\{j 2\pi (R - r\cos(\theta_s-\theta)) / \lambda \right\}}
    {\exp \left\{j 2\pi (R - r\cos\theta) / \lambda \right\}} \mathrm{d}\theta \\
    &= \smallint_0^{2\pi} \exp \left\{j 2\pi r (\cos\theta-\cos(\theta_s-\theta)) / \lambda \right\} \mathrm{d}\theta.
\end{aligned}
\end{equation}
By substituting $\cos\theta-\cos(\theta_s-\theta)$ with $2 \sin\frac{\theta_s}{2} \sin(\frac{\theta_s}{2}-\theta)$, we have $E(\theta_s) = \smallint_0^{2\pi} \exp \left(j m \sin \left(\theta_s / 2-\theta \right) \right) \mathrm{d}\theta$, where $m = \frac{4\pi r}{\lambda} \sin\frac{\theta_s}{2}$.
The integration over a full period leads to
\begin{equation}
    E(\theta_s) = \smallint_0^{2\pi} \exp( j m \sin \theta)\mathrm{d}\theta.
\end{equation}
Based on the definition of the Bessel function of the first kind~\cite{temme1996special}, we arrive at \eqnref{eq:beamshape}.

\vspace{5pt}
\header{Equation \ref{eq:dist_scatter_ante}.}
Given the antenna location $\vec{p}^a_t = (r\cos \omega t + vt\cos\theta_v, r\sin \omega t + vt\sin\theta_v, 0)$ and reflector location $(R_n\cos\theta_n,\allowbreak R_n\sin\theta_n, 0)$, the square of their distance is 
\begin{equation}
    d^2 = [R_n - r\cos(\omega t - \theta_n) - vt\cos(\theta_v-\theta_n)]^2 + Q,
\end{equation}
where $Q=r^2\sin^2(\omega t-\theta_n) + v^2t^2\sin^2(\theta_v-\theta_n)+2rvt\sin(\omega t-\theta_n)\sin(\theta_v-\theta_n)$.
Since $r^2$, $(vt)^2$, and $rvt$ are small compared to $R_n^2$, the term $Q$ can be neglected, resulting in \eqnref{eq:dist_scatter_ante}.

\vspace{5pt}
\header{Equation \ref{eq:linear_d_final}.}
Expanding \eqnref{eq:linear_d} yields $d'(t, t_c) = R_n - vt\cos(\theta_v-\theta_n) -2r \sin\frac{\omega t_c-\theta_n}{2} \sin\frac{\omega t_c-2 \omega t + \theta_n}{2}$. Since reflectors are only observed within the FoV of the radar,
the terms $\omega t_c - \theta_n$, $\omega t_c-\omega t$, and $\theta_n - \omega t$ are relatively small (<30$^\circ$).
With the approximation $\sin x \approx x$, we have 
\begin{equation*}
\begin{aligned}
    d'(t, t_c) &\approx R_n - vt\cos(\theta_v-\theta_n) -\frac{r}{2} (\omega t_c-\theta_n)(\omega t_c-2 \omega t + \theta_n) \\
    &= r\omega t(\omega t_c-\theta_n) - vt\cos(\theta_v-\theta_n) + \mathrm{const.},
\end{aligned}
\end{equation*}
which becomes \eqnref{eq:linear_d_final}.